\input amstex
\magnification 1200
\TagsOnRight
\def\qed{\ifhmode\unskip\nobreak\fi\ifmmode\ifinner\else
 \hskip5pt\fi\fi\hbox{\hskip5pt\vrule width4pt
 height6pt depth1.5pt\hskip1pt}}
\NoBlackBoxes
\baselineskip 19 pt
\parskip 5 pt
\def\stretch {\noalign{\medskip}}
\define \bC {\bold C}
\define \bCp {\bold C^+}
\define \bCm {\bold C^-}
\define \bCpb {\overline{\bold C^+}}
\define \bCmb {\overline{\bold C^-}}
\define \ds {\displaystyle}
\define \bR {\bold R}
\define \bm {\bmatrix}
\define \endbm {\endbmatrix}

\centerline {\bf ON THE BOUND STATES OF}
\vskip -5 pt
\centerline {\bf THE DISCRETE SCHR\"ODINGER EQUATION}
\vskip -5 pt
\centerline {\bf WITH COMPACTLY SUPPORTED POTENTIALS}

\vskip 4 pt
\centerline {Tuncay Aktosun}
\vskip -8 pt
\centerline {Department of Mathematics}
\vskip -8 pt
\centerline {University of Texas at Arlington}
\vskip -8 pt
\centerline {Arlington, TX 76019-0408, USA}
\vskip -8 pt
\centerline {aktosun\@uta.edu}

\vskip 4 pt
\centerline {Abdon E. Choque-Rivero}
\vskip -8 pt
\centerline {Instituto de F\'{\i}sica y Matem\'aticas}
\vskip -8 pt
\centerline {Universidad Michoacana de San Nicol\'as de Hidalgo}
\vskip -8 pt
\centerline {Ciudad Universitaria, C.P. 58048}
\vskip -8 pt
\centerline {Morelia, Michoac\'an, M\'exico}
\vskip -8 pt
\centerline {abdon\@ifm.umich.mx}

\vskip 4 pt
\centerline {Vassilis G. Papanicolaou}
\vskip -8 pt
\centerline {Department of Mathematics}
\vskip -8 pt
\centerline {National Technical University of Athens}
\vskip -8 pt
\centerline {Zografou Campus}
\vskip -8 pt
\centerline {157 80, Athens, Greece}
\vskip -8 pt
\centerline {papanico\@math.ntua.gr}

\noindent {\bf Abstract}:
The discrete Schr\"odinger operator
with the Dirichlet boundary condition is considered on the half-line lattice
$n\in \{1,2,3,\dots\}.$ It is assumed that
the potential belongs to class $\Cal A_b,$ i.e.
it is real valued,
vanishes when $n>b$ with $b$ being a fixed positive integer, and
is nonzero at $n=b.$ The proof is provided
to show that the corresponding number of bound states, $N,$
must satisfy the inequality $0\le N\le b.$
It is shown that for each fixed nonnegative integer $k$
in the set
$\{0,1,2,\dots,b\},$ there exist infinitely many potentials
in class $\Cal A_b$ for which the corresponding Schr\"odinger
operator has exactly $k$ bound states.
Some auxiliary results are presented
to relate the number of
bound states to the number of
real resonances associated
with the corresponding Schr\"odinger operator.
The theory presented is illustrated with some
explicit examples.

\vskip 5 pt
\par \noindent {\bf Mathematics Subject Classification (2010):}
39A70 47B39 81U15 34A33
\vskip -8 pt
\par\noindent {\bf Short title:} Bound states for the discrete Schr\"odinger operator
\vskip -8 pt
\par\noindent {\bf Keywords:} Discrete Schr\"odinger operator,
half-line lattice, bound states, resonances,
\vskip -8 pt
compactly-supported potential,
number of
bound states

\newpage

\noindent {\bf 1. INTRODUCTION}
\vskip 3 pt

We consider the discrete Schr\"odinger equation on the half-line
 lattice, i.e., the difference equation
$$-\psi_{n+1}+2\psi_n-\psi_{n-1}+V_n\,\psi_n=\lambda \,\psi_n, \qquad
   n\ge 1, \tag 1.1$$
with the Dirichlet boundary condition
$$\psi_0=0.\tag 1.2$$
Here, the discrete independent variable $n$ takes positive integer
 values, the boundary of the half-line lattice
 corresponds to $n=0,$
$V_n$ denotes the value of the potential $V$ at the lattice
 location $n,$ $\lambda$ is the spectral parameter, and $\psi_n$
 denotes the value of the wavefunction at the  location $n$. We
 assume that the potential is real valued, i.e.
$$V_n=V_n^\ast, \qquad n\ge 1,\tag 1.3$$
where the asterisk is used for complex conjugation.

There are various physical models [6] governed by the
 discrete Schr\"odinger equation on a half-line lattice,
 and such models describe the
 quantum mechanical behavior of a particle of energy $\lambda$
 in a semi-infinite crystal where the particle experiences at each lattice point
 the
 force associated with the potential value $V_n.$

In this paper we assume that the potential belongs to class
 $\Cal A_b$, which is defined below as in [2].

\noindent {\bf Definition 1.1} {\it The potential $V$ appearing in (1.1) belongs to class
 $\Cal A_b$ if the $V_n$-values are real and the support of the potential
 $V$ is confined to the finite set $\{1,2,\ldots,b\}$ for some
 positive integer $b,$ i.e. $V_n=0$ for $n>b$ and $V_b\ne 0.$}

We refer to a potential $V$ in class $\Cal A_b$ as a compactly-supported
 potential, and we see that $b$ in the definition of $A_b$
 refers to the smallest positive integer beyond which the potential
 vanishes. 
 In class $\Cal A_b,$ it is possible to have
 $V_n=0$ for some or all
 $n$-values with $1\le n<b$ but we must have $V_b\ne 0.$
 Let us remark that the trivial potential where
 $V_n=0$ for all $n\ge 1$ can either be included in class $A_b$
 by letting $b$ also take the value $b=0$
 or that trivial potential can be studied separately.

 When the potential $V$ belongs to class $A_b$ the discrete
 Schr\"odinger operator corresponding to (1.1) with the
 Dirichlet boundary condition (1.2) is a selfadjoint operator
 acting on square-summable functions
 on the half-line lattice
 and its spectrum is well understood [1-5].
The corresponding spectrum has two parts, where the first part is the
 continuous spectrum $\lambda\in [0,4]$, and the second part is the
 discrete spectrum consisting of a finite number $\lambda$-values
 in the set $\lambda\in(-\infty,0)\cup (4,+\infty)$.
 Each $\lambda$-value in the interval $(0,4)$ corresponds to a
 scattering state, and each $\lambda$ in the discrete spectrum
 corresponds to a bound state, and the values $\lambda=0$ and $\lambda=4$
 correspond to the edges of the continuous spectrum.

We denote the number of discrete eigenvalues by $N.$ In this paper
 we prove that, when the potential $V$ belongs to class $\Cal A_b$ for some fixed
 positive integer $b,$
 the value of $N$ is restricted to the set $\{0,1,\dots, b\},$
 where every value in the set including $N=0$ and $N=b$ is always attained by
  an infinite number of potentials in class $\Cal A_b.$
 Thus, in our paper we prove two main results for potentials
  in class $\Cal A_b.$ The first is that $0\leq N\leq b$ for any potential
in class $\Cal A_b$ having $N$ bound states. The second is that for every
  integer $k$ in the set $\{0,1,\dots, b\},$ there exists at least one
 potential in class $\Cal A_b$ for which there are exactly
 $k$ bound states, and in fact there are infinitely many
 potentials in $\Cal A_b$ for which there are exactly $k$ bound states.

Let us use $N_-$ to denote the number of bound states located in
  the interval $\lambda\in (-\infty,0)$ and use $N_+$ to denote
  the number of bound states located in the interval
   $\lambda\in(4,+\infty).$
While proving the two aforementioned main results, we also obtain
   some upper bounds on each of $N_-$ and $N_+,$ where
 the bounds are related to certain integers related to
 the number of certain real resonances associated with the
 Schr\"odinger operator for (1.1) and (1.2).

Our paper is organized as follows. In Section~2 we briefly
  present the preliminaries needed to prove our two main results.
  This involves the introduction of the Jost solution $f_n$
  to (1.1), the Jost function $f_0$ associated with (1.1)
   and (1.2),
    the alternate spectral parameter $z$ related to $\lambda$
    as in (2.1), and the real and complex resonances associated
    with (1.1) and (1.2). In Section~3 we prove our
    two main results; namely, we have $0\le N\le b$
    for every potential in class $\Cal A_b$
    and that $\Cal A_b$ contains infinitely many
    potentials
    with $k$ bound states for every $k\in\{0,1,\dots,b\}.$
    We also obtain certain upper bounds on $N_-$ and on $N_+.$
         Finally, in Section~4 we present various explicit
     examples illustrating the theory presented in Section~3. In
     particular, we illustrate the inequalities $0\le N\le b$  and the attainment
     of $N$ being equal to any integer
     between zero and $b$ in class $\Cal A_b.$

\vskip 10 pt
\noindent {\bf 2. PRELIMINARIES}
\vskip 3 pt

For the analysis of bound states of the discrete Schr\"odinger
     operator corresponding to (1.1) and (1.2), it is
     useful to use the parameter $z$ related
     to the spectral parameter $\lambda$ as
$$z=1-\frac{\lambda}{2}+\frac{1}{2}\sqrt{\lambda(\lambda-4)},\tag 2.1$$
      where the square root denotes the principal branch of the
      complex square-root function.
            Under the transformation $\lambda\mapsto z$
      specified in (2.1),
      the extended real $\lambda$-axis is mapped in a one-to-one manner onto the
      boundary of the upper half of the unit disk
      in the complex-$z$ plane. In
      particular, the interval $\lambda\in (-\infty,0)$ is mapped
      to the interval $z\in (0,1)$, the interval $\lambda\in (0,4)$
      is mapped to the upper
      semicircle $z=e^{i\theta}$ with $0<\theta<\pi$, and the
      interval $\lambda\in (4,+\infty)$ is mapped to the interval
      $z\in (-1,0)$.
       Using (2.1) we can transform (1.1) into
$$-\psi_{n+1}+2\,\psi_n-\psi_{n-1}=\left(z+z^{-1}+V_n\right)\psi_n, \qquad
   n\ge 1.\tag 2.2$$

A particular solution to (2.2), whose values are denoted by $f_n,$
  with the asymptotics
$$f_n=z^n[1+o(1)], \qquad
   n\to \infty, \tag 2.3$$
  is usually called the Jost solution.
  We occasionally also use $f_n(z)$ to denote $f_n$.
   The value $f_0$, i.e. the value of $f_n$ evaluated at $n=0$,
   is known as the Jost function. When the potential $V$ belongs
   to class $\Cal A_b$, as seen from (2.2) and (2.3)
   we have
$$f_{n}=z^n, \qquad n\ge b, \tag 2.4$$
$$\cases
   f_{b-1}=-f_{b+1}+(z+z^{-1}+V_b)f_b, \\
   \stretch
   f_{b-2}=-f_{b}+(z+z^{-1}+V_{b-1})f_{b-1}, \\
     \stretch
   f_{b-3}=-f_{b-1}+(z+z^{-1}+V_{b-2})f_{b-2}, \\
  \qquad \vdots\\
   f_{1}=-f_{3}+(z+z^{-1}+V_2)f_2, \\
     \stretch
   f_{0}=-f_{2}+(z+z^{-1}+V_1)f_1.\endcases
\tag 2.5$$

A bound-state solution corresponds to a square-summable solution
 to (1.1) satisfying the Dirichlet boundary condition
 (1.2).

In the following theorem we summarize some basic facts needed
 later on.

\noindent {\bf Theorem 2.1} {\it Assume that
the potential $V$ belongs to class $\Cal A_b$ specified in
 Definition~1.1. We then have the following:}

\item{(a)} {\it The Jost function $f_0(z)$ associated with (1.1)
  is a polynomial in $z$ of degree $2b-1$ and it has the form
$$f_{0}(z)=1+K_{01}z+K_{02}z^2+\cdots+ K_{0(2b-1)}z^{2b-1},
    \tag 2.6$$
  where each double-indexed coefficient $K_{0j}$ is real valued
  and we have $K_{0(2b-1)}=V_b.$}

\item{(b)} {\it Each coefficient $K_{0j}$ in (2.6)
  is a polynomial in the multivariable $(V_1,V_2,\dots,V_b).$
In each term in the coefficient $K_{0j},$ each $V_n$-value
for $n=1,\dots,b$ appears either to the first power
or does not appear at all.
  The Jost function $f_0(z)$, viewed as a polynomial in
   $(V_1,V_2,\dots, V_b)$ contains a single monomial
 with degree $b,$ and that monomial is given
   by $\left(V_1\,V_2\cdots V_b\right) z^b$. Any other term in $f_0(z)$
   has a degree of $b-1$ or less. Thus, the Jost function $f_0(z)$
   has the unique decomposition}
$$f_{0}(z)=F(z)+G(z), \tag 2.7$$
{\it where we have defined}
$$F(z):=\left(\ds\prod_{j=1}^b V_j\right)z^b, \tag 2.8$$
$$G(z):=f_0(z)-\left(\ds\prod_{j=1}^b V_j\right)z^b. \tag 2.9$$

\item{(c)} {\it For each $b\ge 1,$ there are exactly $2b-1$ zeros
  of $f_0(z)$ in the complex-$z$ plane. Such zeros occur either when $z\in \bR\setminus \{0\}$
  or they occur as complex-conjugate pairs symmetrically located
  with respect to the real axis $\bR$ in the complex-$z$ plane. The point
  $z=0$ does not correspond to a zero of $f_0(z).$}

\item{(d)} {\it The zeros of $f_0(z)$ occurring in $z\in (-1,0)\cup(0,1)$
  are each simple, and each such zero corresponds to a bound state
  of
  the discrete Schr\"odinger operator associated with (1.1) and (1.2).
  We use $N_{-}$ to denote the number of zeros of $f_0(z)$ in the
  interval $z\in (-1,0)$, use $N_{+}$ to denote the number of zeros
  of $f_0(z)$ in the interval $z\in (0,1)$, and use $N$ to denote
  the number of zeros of $f_0(z)$ when $z\in (-1,0)\cup(0,1)$. Thus, we have}
$$N=N_-+N_+.\tag 2.10$$

\item{(e)} {\it The Jost function $f_0(z)$ may have a simple zero at $z=-1$
  and may have a simple zero at $z=1$. Such zeros do not
  correspond to bound states for the  discrete Schr\"odinger
  operator. Let us use $\mu_-$ to denote the number of zeros of
  $f_0(z)$ at $z=-1$ and use $\mu_+$ for the number of zeros of
  $f_0(z)$ at $z=1.$ Hence, we have}
 $$
  \mu_-=\cases 1,\qquad  f_0(-1)=0, \\
    \stretch
            0,\qquad f_0(-1)\ne 0,\endcases\tag 2.11$$
$$\mu_+=\cases 1,\qquad  f_0(1)=0, \\
  \stretch
            0,\qquad f_0(1)\ne 0,\endcases\tag 2.12$$

\item{(f)} {\it The zeros of $f_0(z)$ when $z\in (-\infty,-1)$ are not
  necessarily simple. Similarly, the zeros of $f_0(z)$ when $z\in (1,+\infty)$
  are not necessarily simple. We refer to the zeros of $f_0(z)$
  when $z\in(-\infty,-1)\cup(1,+\infty)$
  as real resonances for the discrete Schr\"odinger operator.}

\item{(g)} {\it The nonreal zeros of $f_0(z)$ cannot occur
  inside or on the unit circle $|z|=1$. Such zeros,
  if they exist, are not necessarily simple and they occur
  as complex-conjugate pairs located outside the unit circle
  in the complex-$z$ plane.}

\item{(h)} {\it We have}
$$2b-1=Z(-\infty,1]+ Z(-1,0)+Z(0,1)+Z[1,+\infty)+2\,Z_c,\tag 2.13$$
{\it where the nonnegative integer
  $Z(-\infty,1]$ is the number of zeros of $f_0(z)$ in the interval
  $z\in(-\infty,-1],$ the nonnegative integer
  $Z(-1,0)$ is the number of zeros of $f_0(z)$ in the interval
  $z\in(-1,0),$
  the nonnegative integer
  $Z(0,1)$ is the number of zeros of $f_0(z)$ in the interval
  $z\in(0,1),$
  the nonnegative integer
  $Z[1,+\infty)$ is the number of zeros of $f_0(z)$ in the interval
  $z\in[1,+\infty),$
  and the nonnegative integer $Z_c$
  is the number of zeros of $f_0(z)$ located in the interior of the
  upper-half complex-$z$ plane.}

\noindent PROOF: For the results stated in (a), (c), (d), (e) we refer the
reader to Theorems~2.2, 2.4, 2.5 of [1].
 The proof of (b) is directly obtained from (2.5) as follows.
 From (2.50) of [2] we see that each coefficient $K_{0j}$ in (2.6)
 is a polynomial in $(V_1,V_2,\dots, V_b)$ and that $f_0(z)$ has the
 form
 $$
 f_0(z)=1+\left(V_1+\cdots+V_b\right)\,z+\cdots+V_b\left(V_1+\cdots+V_{b-1}
 \right)z^{2b-2}+
 V_b\,z^{2b-1}. \tag 2.14$$
In each line of (2.5) expressing
$f_{n-1}$ for $n=1,\dots,b,$ only one single potential
value, i.e. $V_n$ appears. Thus, in expressing
$f_0(z)$ as in (2.14), each term in $f_0(z)$ contains
$V_n$ either to the first power or to the zeroth
power.
Let us view each $f_n$ as a polynomial in the multivariable
   $(V_1,\ldots,V_b)$. From (2.5) we observe that the
   highest-degree term in $f_{b-1}$ is the single monomial given
   by $V_b\,f_b$ or equivalently by $V_b\,z^b;$ the highest-degree term
   in $f_{b-2}$ is the term  $V_{b-1}f_{b-1}$
   or equivalently $V_{b-1}V_b\,z^b;$ the highest-degree term
   in $f_{b-3}$ is the term $V_{b-2}f_{b-2}.$
   Continuing in this manner, from (2.5) we see that the
   highest degree term in $f_0$ is the single monomial given by
   $(V_1V_2\cdots V_b)z^b$. Thus, the proof of (b) is complete.
   For (f) we refer the reader to (4.13) and (4.14) in
   Example~4.3 where we
   illustrate a double
   zero of $f_0(z)$ when $z\in(-\infty,-1)$ and to
   (4.11) and (4.12) in Example~4.3 for a double
   zero in $z\in(1,+\infty).$ For (g) we refer the
   reader to Theorem~2.4 of [1] and to (4.18) and (4.19) in
   Example~4.4 where
   we illustrate a double complex zero.
Finally, the proof of (h) is obtained as follows. By (c) we know that
$f_0(z)$ is a polynomial in $z$ of degree $2b-1$ and that it has exactly
 $2b-1$ zeros in the complex-$z$ plane. By (c) we also know that the
 zeros of $f_0(z)$ off the real axis must occur in complex conjugate
 pairs and hence the number of such zeros can be represented as
 $2\,Z_c,$ where $Z_c$ is the number of zeros of $f_0(z)$ in $z\in
 \bCp.$
  From (c) we also know that $f_0(z)$ cannot have a zero at $z=0.$
  Thus, (2.13) holds, which expresses the total number of zeros
  of $f_0(z)$ in terms of the number of zeros at various different locations in
  the complex-$z$ plane. \qed

By Theorem~2.1(d) we know that each zero of $f_0(z)$ when
$z\in(-1,0)\cup (0,1)$ corresponds to a bound state of the discrete
Schr\"odinger operator associated with (1.1) and (1.2).
 It is known [2] that for each bound-state zero, there exists a positive
 constant, known as the Marchenko norming constant, and when the
 potential belongs to class $\Cal A_b,$ the Marchenko norming constant $c_k$
 corresponding to the zero $\alpha_k$ of $f_0(z)$ with
  $\alpha_k\in (-1,0)\cup(0,1)$ is related to the Jost function
  $f_0(z)$ as
$$c_k^2=\text{Res}\left[\ds\frac{f_0\left(\ds\frac{1}{z}\right)}{z\,f_0(z)},\alpha_k\right]
,\tag 2.15$$
  where the notation
  $\text{Res}[h(z),\alpha]$ is used to denote the residue of the function
  $h(z)$ at $z=\alpha.$. The expression given in (2.15)
  follows from (2.45), (2.49) and (3.19) of [2].

In the next proposition, we express the right-hand side of (2.15)
  in terms of all the zeros of $f_0(z).$

\noindent {\bf Proposition 2.2} {\it Assume that
the potential $V$ appearing in (1.1) belongs to class $\Cal A_b$ specified in
 Definition~1.1. Then, the Marchenko
 norming constant $c_k$ appearing in (2.15) satisfies}
$$c_k^2=\ds\frac{1}{\alpha_k^{2b}}\,\ds
    \frac{\ds\prod_{s=1}^{2b-1}\left(1-\alpha_k\,\alpha_j\right)}
   {\ds\prod_{j\neq k}\left(\alpha_k-\alpha_j\right)},\tag 2.16$$
{\it  where $\{\alpha_j\}_{j=1}^{2b-1}$ is the set of all zeros of
  $f_0(z)$ and the product appearing in the denominator of (2.16)
  is over all $j=1,\dots, 2b-1$ except $j\ne k.$}

\noindent PROOF: From Theorem~2.1(a) we know that $f_0(z)$ is a polynomial in
  $z$ of degree $2b-1,$ and hence it can be written in terms of its
  zeros as
$$f_0(z)=\left(1-\ds\frac{z}{\alpha_1}\right)\left(1-\ds\frac{z}{\alpha_2}\right)
   \cdots\left(1-\ds\frac{z}{\alpha_{2b-1}}\right),
\tag 2.17$$
  where by Theorem~2.1(d) we know that those $\alpha_k$ values
  confined to $(-1,0)\cup(0,1)$ are simple.
  We can write $f_0(z)$ in terms of its zeros also as
$$f_0(z)=V_b\,(z-\alpha_1)\,(z-\alpha_2)
   \cdots (z-\alpha_{2b-1}),
\tag 2.18$$
  where we have used the fact that the coefficient of
  $z^{2b-1}$ in the expression (2.14) is equal to
  $V_b$.
  From (2.46), (2.49), (3.19), and (3.22) of [2], we have
$$c_k^2=
   \ds \frac{f_0\left(\ds\frac{1}{\alpha_k}\right)}{\alpha_k\, \dot f_0(\alpha_k)}
\tag 2.19$$
  where $\dot f_0(\alpha_k)$ is $df_0/dz$ evaluated at
  $z=\alpha_k$.
  From (2.18) we get
$$f_0(1/\alpha_k)=V_b\,\ds \prod_{s=1}^{2b-1}
  \left(\ds\frac{1}{\alpha_k}-\alpha_s\right),\tag 2.20$$
  $$\dot f_0(\alpha_k)=V_b\,\ds \prod_{j\ne k}(\alpha_k-\alpha_j).
\tag 2.21$$
  Using (2.20) and (2.21) in (2.19) we get
$$c_k^2=
    \ds\frac{\ds\prod_{l=1}^{2b-1}\left(\ds\frac{1}{\alpha_k}-\alpha_l\right)}
   {\alpha_k\, \ds\prod_{j\ne k}(\alpha_k-\alpha_j)}
,$$
  or equivalently
$$c_k^2=\ds\frac{1}{\alpha_k}\,\ds\frac{1}{\alpha_k^{2b-1}}\,\ds
    \frac{\ds\prod_{l=1}^{2b-1}
    (1-\alpha_k\,\alpha_l)}
   {\ds\prod_{j\ne k}(\alpha_k-\alpha_j)},
 $$
 which yields (2.16). \qed

\vskip 10 pt
\noindent {\bf 3. MAIN RESULTS}
\vskip 3 pt

In order to prove our two main results, i.e. the number of
  bound states $N$ for any potential in class $\Cal A_b$ must satisfy
$0\leq N\le b$
  and that every integer $k$ in the set $\{0,1,\dots,b\}$
  is equal to the number of bound states for
at least one potential, and in fact infinitely many potentials, in class $\Cal A_b,$
  we first need to prove some auxiliary results.

  In reference to the nonzero integers
  appearing in (2.13), let us define
$$p:=Z(-\infty,-1], \tag 3.1$$
$$q:=Z(-\infty,-1]+Z(-1,0), \tag 3.2$$
$$r:=Z(-\infty,-1]+Z(-1,0)+Z(0,1), \tag 3.3$$
$$s:=Z(-\infty,-1]+Z(-1,0)+Z(0,1)+Z[1,+\infty).\tag 3.4$$
The integers $p,$ $q,$ $r,$ $s$
 help us to order the real zeros of $f_0(z)$ in an increasing order
 and they indicate the
 location of those zeros as
$$\alpha_1, \alpha_2, \dots, \alpha_p \in (-\infty,-1],\tag 3.5$$
$$\alpha_{p+1}, \alpha_{p+2}, \dots, \alpha_q \in (-1,0), \tag 3.6$$
$$\alpha_{q+1}, \alpha_{q+2}, \dots, \alpha_r \in (0,1),\tag 3.7$$
$$\alpha_{r+1}, \alpha_{r+2}, \dots, \alpha_{s}
   \in [1,+\infty),\tag 3.8$$
so that $\alpha_{s+1},$ $\alpha_{s+2},\dots, \alpha_{2b-1}$
correspond to the
  nonreal zeros of $f_0(z).$ We know from Theorem~2.1(g) that those nonreal
  zeros  must occur in complex-conjugate pairs and hence we have
$$2b-1-s=2\,Z_c, \tag 3.9$$
  where $Z_c$ is the nonnegative integer appearing in (2.13).

  Let us use $\text{sgn}$ to denote the signature function, i.e.
   $$
  \text{sgn}(x):=\cases
1,\qquad  x>0,\\
\stretch
-1,\qquad x<0 .\endcases
 $$

  In the following proposition we consider the sign of the product
appearing in the numerator of (2.16).

  \noindent {\bf Proposition 3.1} {\it Assume that
the potential $V$ appearing in (1.1) belongs to class $\Cal A_b,$
and further assume that there exists at least one bound state.
Let $\{\alpha_j\}_{j=1}^{2b-1}$ denote the
set of zeros of $f_0(z)$ ordered as indicated in (3.5)-(3.8).
Then, for the bound state occurring at $z=\alpha_k$ we have}
$$
\text{\rm sgn}\left(\ds\prod_{j=1}^{2b-1}(1-\alpha_k\,
  \alpha_j)\right)=\cases
\text{\rm sgn}\left(
\ds\prod_{j=1}^{p}(1-\alpha_k\,
  \alpha_j)\right),\qquad \alpha_k\in(-1,0), \\
  \stretch
\text{\rm sgn}\left(
\ds\prod_{j=r+1}^{s}(1-\alpha_k\,
  \alpha_j)\right),\qquad \alpha_k\in(0,1),\endcases
\tag 3.10$$
{\it where $p,$ $r,$ and $s$ are the integers defined in (3.1), (3.3),
 and (3.4), respectively.}

\noindent PROOF:
Let us omit $(1-\alpha_k\,\alpha_j)$ from the argument of the
 product, and write e.g. $\prod_1^p$ to denote
 $\prod_{j=1}^p (1-\alpha_k\,\alpha_j)$. We have
$$
\ds\prod_1^{2b-1}=\left( \ds\prod_1^p\right) \left( \prod_{p+1}^q\right)\left(
\ds\prod_{q+1}^r \right)\left( \ds \prod_{r+1}^s\right)\left( \ds \prod_{s+1}^{2b-1}\right).
\tag 3.11$$
With the help of (3.6) and (3.7) we see that each
 of $\prod_{p+1}^q$ and $\prod_{q+1}^r$ is positive
 because $|\alpha_j|<1$ when $p+1\le j\le r.$
 By Theorem~2.1(g), the zeros $\alpha_j$ for $s+1\le j\le 2b-1$
 occur in complex-conjugate pairs and hence
 the quantity $\prod_{s+1}^{2b-1}$ appearing in (3.11)
 consists of the products of the form $|1-\alpha_k\,\alpha_j|^2$ when the
 $\alpha_j$-values consist of all nonreal zeros of $f_0(z)$ in the
 upper-half complex-$z$ plane. Since $\alpha_k$ is real and such
 $\alpha_j$-values are nonreal, we then conclude that $\prod_{s+1}^{2b-1}$
 is positive. Let us now consider the sign of $\prod_1^p$
 appearing in (3.11). As seen from (3.5), (3.6),
 and (3.7) for $1\le j\le p$ we have $1-\alpha_k\,\alpha_j>0$
 if $\alpha_k\in(0,1).$ Let us also consider the sign of $\prod_{r+1}^s$
 appearing in (3.11). As seen from (3.6), (3.7),
 and (3.8), for $r+1\leq j\leq s$, we have $1-\alpha_k\,\alpha_j>0$
 if $\alpha_k\in (-1,0)$. Thus, from (3.11) we directly conclude
 (3.10). \qed

In the following proposition, we consider the sign of the product
appearing in the denominator of (2.16).

\noindent {\bf Proposition 3.2} {\it Assume that
the potential $V$ appearing in (1.1) belongs to class $\Cal A_b,$
and further assume that there exists at least one bound state.
Let $\{\alpha_j\}_{j=1}^{2b-1}$ denote the
set of zeros of $f_0(z)$ ordered as indicated in (3.5)-(3.8).
Then, for the bound state occurring at $z=\alpha_k$ we have}
$$
 \text{\rm sgn}\left(\ds\prod_{j\ne k}(\alpha_k-\alpha_j)
 \right)=(-1)^{k-1}, \qquad k=p+1,\ldots, r, \tag 3.12
$$
{\it where $p$ and $r$ are the respective integers appearing in (3.1)
 and (3.3), and the product $\prod_{j\ne k}$ is over
 all $j$-values with $1\le j\le 2b-1$ except $j=k,$ and {\rm sgn}
 denotes the signature function.}

\noindent PROOF:
  Let us drop $(\alpha_k-\alpha_j)$ from the argument of the
  product and use
 $\prod_{j\ne k}$ to denote the product on the left-hand
 side of (3.12). We have
$$\ds\prod_{j\ne k}=\left(\ds\prod_1^{k-1}\right)\left(
 \ds\prod_{k+1}^s\right)\left( \ds\prod_{s+1}^{2b-1}\right),\tag 3.13$$
 where $\prod_1^{k-1}$ denotes
  $\prod_{j=1}^{k-1}(\alpha_k-\alpha_j),$ etc. From
  (3.5)-(3.8) we see that $\prod_1^{k-1}$
   is positive because $\alpha_k>\alpha_j$ when $1\le j<k.$
   The sign of $\prod_{k+1}^s$ is the same as the sign of
   $(-1)^{s-k}$ because each factor $(\alpha_k-\alpha_j)$
   is negative as we
   have $\alpha_k<\alpha_j$ for $k+1\le j\le s.$ Furthermore, the
   quantity $\prod_{s+1}^{2b-1}$ is positive because
   it consists of the products $(\alpha_k-\alpha_j)(\alpha_k-\alpha_j^\ast),$
which is equivalent to $|\alpha_k-\alpha_j|^2,$ when the $\alpha_j$-values are on the
   upper-half complex-$z$ plane, as indicated in Theorem~2.1(g).
   Since $\alpha_k$ is real and such $\alpha_j$-values area nonreal,
   each factor $|\alpha_k-\alpha_j|^2$ is positive and hence
   $\prod_{s+1}^{2b-1}$ is positive.
   Thus, from (3.13) we conclude that the sign of the
   right-hand side is the same as the sign of $(-1)^{s-k}.$
   On the other hand, from (2.13) and (3.9) we know that $s$
   is an odd positive integer. Thus the sign of $(-1)^{s-k}$ is the
   same as the sign of $(-1)^{k-1}$, which establishes (3.12). \qed

We recall from Theorem~2.1(f) that we refer to the real zeros of
$f_0(z)$ when $z\in (-\infty,-1)\cup(1,+\infty)$ as the real resonances
whereas the zeros when $z\in(-1,0)\cup (0,1)$ correspond to the
bound states.
In the following proposition we investigate the relationship
between the number of real resonances and the number of bound states
for a potential in class $\Cal A_b.$

\noindent {\bf Proposition 3.3} {\it Assume that
the potential $V$ appearing in (1.1) belongs to class $\Cal A_b.$
As in Theorem~2.1(h), let $Z(-\infty,-1],$ $Z(-1,0),$ $Z(0,1),$ and
$Z[1,+\infty)$ denote the number of zeros
of $f_0(z)$ in the intervals $(-\infty,-1],$
$(-1,0),$ $(0,1),$ and $[0,+\infty),$ respectively.
Furthermore, assume that there exists at least one bound state.
Let $\{\alpha_j\}_{j=1}^{2b-1}$
denote the number of zeros of $f_0(z)$ in the intervals
 $(-\infty,-1],$ $(-1,0),$ $(0,1),$ and $[1,+\infty),$ respectively.
 We then have the following:}

\item{(a)} {\it If $Z(-1,0)=0$ then}
$$Z(-\infty,-1]= Z(-1,0)-1+\varepsilon_{-},\tag 3.14$$
{\it where $\varepsilon_-$ is a positive integer with
    $\varepsilon_-\ge 1.$}

  \item{(b)} {\it If $Z(-1,0)\ge 1$ then}
$$Z(-\infty,-1]= Z(-1,0)-1+\varepsilon_{-},\tag 3.15$$
{\it where $\varepsilon_-$ is a nonnegative integer with
    $\varepsilon_{-}\ge 0.$}

 \item{(c)} {\it If $Z(0,1)=0$ then}
$$Z[1,+\infty)= Z(0,1)-1+\varepsilon_+,\tag 3.16$$
{\it where $\varepsilon_+$ is a positive integer with
    $\varepsilon_+\ge 1.$}

\item{(d)} {\it If $Z(0,1)\ge 1$ then}
$$Z[1,+\infty)= Z(0,1)-1+\varepsilon_+,\tag 3.17$$
{\it where $\varepsilon_+$ is a nonnegative integer with
    $\varepsilon_+\ge 0.$}

    \item{(e)} {\it We have}
$$Z(-\infty,-1]\ge\cases
     0,\qquad Z(-1,0)=0,\\
 Z(-1,0)-1,\qquad Z(-1,0)\ge 1,\endcases\tag 3.18$$
$$Z[1,\infty)\ge\cases
0,\qquad Z(1,0)=0,\\
 Z(1,0)-1,\qquad Z(1,0)\ge 1.\endcases\tag 3.19$$

\noindent PROOF:
Note that (3.14) and (3.16) automatically follow from
  the facts that $Z(-\infty,-1]$ and $Z[1,+\infty)$
  must both be nonnegative as they represent the number
  of zeros of $f_0(z)$ in the appropriate intervals. So,
  it is enough to prove (3.15) and (3.17) only.
  If $Z(-1,0)\ge 1$ then there exists at least one $\alpha_k$-value
  in the interval $z\in(-1,0),$ which corresponds to a bound state.
  Let $c_k$ be the corresponding Marchenko norming constant. From
  (2.16), (3.12), and the first line of (3.10)
  we get
$$\text{\rm sgn} \left(c_k^2\right)=(-1)^{k-1} \,\text{\rm sgn}\left(\ds\prod_{j=1}^p
  (1-\alpha_k\,\alpha_j) \right),
  \tag 3.20$$
where $p$ is the integer defined in (3.1).
  Being a norming constant, we have $c_k>0$ and hence from (3.1)
  we conclude that
$$\text{\rm sgn} \left(\ds\prod_{j=1}^p
  (1-\alpha_k\,\alpha_j)\right)=(-1)^{k-1}, \qquad k=p+1,\dots,q,\tag 3.21$$
where $q$ is the integer appearing in (3.2) and
  (3.6). Defining
$$P_-(z):=\ds\prod_{j=1}^p (1-\alpha_j\,z),\tag 3.22$$
let us investigate the sign of $P_-(z)$ when $z$ takes values in
  the interval $z\in(-1,0).$
  From (3.20) and (3.22) we know that
$$\text{\rm sgn}\left( P_{-}(\alpha_k)\right)=(-1)^{k-1},\qquad k=p+1,\dots, q,
\tag 3.23$$
where $q$ is the integer appearing in (3.2) and (3.6).
 Thus, from (3.23) we conclude that in the interval
 $z\in(-1,0)$, the polynomial $P_-(z)$ defined in (3.22)
 changes sign at least $(q-p-1)$ times.
 This indicates that the polynomial $P_{-}(z)$ must have degree no
 less than $q-p-1.$
 From (3.1) and (3.2) we have $q-p=Z(-1,0),$ and
 from (3.22) we know that the degree of $P_-(z)$ is the
 same as $p,$ which is equal to $Z(-\infty,-1]$ as seen from
 (3.1).
  Thus, we have proved that
$$Z(-\infty,-1)\ge Z(-1,0)-1,\tag 3.24$$
when $Z(-1,0)\ge 1.$
We can then write (3.24) as (3.18), and hence the proof
of (b) is complete.
In a similar way, let us prove (d). If $Z(0,1)\ge 1$ then there
exists at least one $\alpha_k$-value in the interval $z\in(0,1),$
which corresponds to a bound state. Let $c_k$ be the corresponding
Marchenko norming constant. From (2.16), (3.12), and the
second line of (3.10), we get
$$\text{\rm sgn} \left(c_k^2\right)=(-1)^{k-1}\,\text{\rm sgn}\left(\prod_{j=r+1}^s
 (1-\alpha_k\,\alpha_j)\right),\tag 3.25$$
  where $r$ and $s$ are the integers appearing in (3.3),
  (3.4), (3.7), and (3.8). The norming constant
  $c_k$ is positive and hence from (3.25) we conclude that
$$\text{\rm sgn} \left(\prod_{j=r+1}^s
 (1-\alpha_k\alpha_j)\right)=(-1)^{k-1},
 \qquad k=q+1,\ldots, r,\tag 3.26$$
where $q$ is the integer appearing in (3.2), (3.6),
  and (3.7).
Letting
$$P_+(z):=\ds\prod_{j=r+1}^s (1-\alpha_j\,z),\tag 3.27$$
we notice that the degree of $P_+(z)$ is $s-r,$ which is equal
  to $Z[1,+\infty)$ as seen from (3.3) and (3.4).
   Let us investigate the sign of $P_+(z)$ when $z$ takes values
   in the interval $z\in(0,1).$ From (3.36) and (3.27)
   we see that
$$\text{\rm sgn} \left(P_{+}(\alpha_k)\right)=(-1)^{k-1},\qquad k=p+1,\dots, r.
\tag 3.28$$
 Thus, from (3.28) we conclude that in the interval $z\in(0,1)$
 the polynomial $P_+(z)$ changes sign at least $(r-q-1)$ times.
 Hence, the degree of $P_+(z)$ cannot be less than $(r-q-1).$ From
 (3.2) and (3.3) we see that $r-q=Z(0,1),$ and we have
 already seen that the degree of $P_+(z)$ is equal to
 $Z[1,+\infty).$ Thus, we have proved that
$$Z[1,+\infty)\geq Z(0,1)-1,\tag 3.29$$
when $Z(0,1)\ge 1.$ We can write (3.29) in the form
 given in (3.17), and hence the proof of (d) is complete.
  The result in (e) directly follows from (a)-(d).
\qed

In order to prepare the proof of one of our two main theorems, namely that for
 any potential in class $\Cal A_b$ having $N$ bound states we must
 have $0\le N\le b,$ we first prove that for any $b\ge 1$
 there exist infinitely many nontrivial potentials in class $A_b$ for which $N=0$ and also
 there exist infinitely many
 potentials in class $A_b$ for which $N=b.$ We then prove our other main theorem,
 namely, that for each
 $k$ in the set $\{0,1,\dots,b\}$ there are infinitely many potentials
in class $\Cal A_b$ having exactly $k$ bound states.
We remark that in the limiting case where $b=0$ the potential class
$\Cal A_b$ contains only the trivial potential
where $V_n\equiv 0,$ for which we already know
 that $N=0.$ Thus, our main result $0\le N\le b$
 automatically and trivially holds
 also when $b=0.$

In the next theorem we prove that for any fixed $b\ge 1$ the class
 $\Cal A_b$ contains infinitely many potentials with
$N=0.$ We recall that for each potential in class $\Cal A_b$
we have $V_b\ne 0.$

\noindent {\bf Theorem 3.4} {\it For any fixed $b$
with $b\ge 1,$ the class $\Cal A_b$ specified in Definition~1.1
contains infinitely many potentials having no bound
states.}

\noindent PROOF: From Theorem~2.1(b) we already know that
 the Jost function $f_0(z)$ is a polynomial in the multivariable
  $(V_1,V_2,\dots,V_b).$ In fact, $f_0(z)$ considered as such
 a polynomial consists of terms where the degree of each term is
  between $0$ and $b.$ In Theorem~2.1(b) we have seen that the term
  with the largest degree is the unique monomial
given by $(V_1\cdots V_b)z^b.$
  In fact, $f_0(z)$ as a polynomial in $(V_1,\ldots, V_b)$ contains
  a single monomial term with zero degree and that term is the
  term given by $1$ in $f_0(z).$ We already know that the zero potential
   $V_n\equiv 0$ corresponds to $N=0$ with the corresponding Jost function
   being $f_0\equiv 1.$
   By using a small perturbation on the trivial potential
   appropriately, we can get a nontrivial potential with $N=0.$
   This can be seen as follows. In (2.6) every term in $f_0(z)-1,$ viewed as
   a polynomial in $(V_1,\dots, V_b),$ has degree at least one.
Thus, we can choose each
    $|V_j|$ small enough and with $V_b\ne 0$ so that the corresponding $K_{0j}$ appearing
    in (2.6) satisfies
$$|K_{0j}|<\ds\frac{1}{2b},  \qquad 1\le j\le 2b-1.\tag 3.30$$
    By Theorem~2.1(a) we know that each $K_{0j}$ is real and hence
    (3.30) implies that when $z\in (-1,1)$ we get
$$-\ds\frac{1}{2b}<K_{0j}\,z<\ds\frac{1}{2b},  \qquad 1\le j\leq 2b-1.
\tag 3.31$$
 From (2.6) we have the standard inequality
$$|f_0(z)|\ge\left|1-\left|K_{01}\,z+K_{0,2}\,z^2+\cdots
  +K_{0(2b-1)}z^{2b-1}\right|\right|.\tag 3.32$$
Using (3.31) in (3.32), for $z\in(-1,1)$ we obtain
$$|f_0(z)|\ge 1-(2b-1)\ds\frac{1}{2b},\tag 3.33$$
or equivalently
$$|f_0(z)|\ge \ds\frac{1}{2b},\tag 3.34$$
implying that there cannot be any $z$-value in the
interval
$z\in(-1,1)$
 for which $f_0(z)=0.$ Thus, for such potentials we must have $N=0.$
 \qed

In the next theorem we prove that for any fixed $b\ge 1$, the class
 $A_b$ contains infinitely many potentials for which the number
  of bound states
 is exactly $b.$

\noindent {\bf Theorem 3.5} {\it For any fixed $b$
with $b\ge 1,$ the class $\Cal A_b$ specified in Definition~1.1
contains infinitely many potentials having exactly $b$ bound
states.}

\noindent PROOF: From Theorem~2.1(b) we know that we can write the Jost function
$f_0(z)$ as in (2.7) as the sum of $F(z)$ and $G(z)$ where
$F(z)$ is the monomial in $(V_1,\dots, V_b)$ defined in (2.8)
and $G(z)$ is the polynomial in $(V_1,\dots, V_b)$ of degree no larger
than $b-1$ given in (2.9). Let us choose our potential $V$ in class
$\Cal A_b$ such
that $|V_b|>1$ and
$$
|V_j|=|V_b|, \qquad j=1,\dots, b-1.\tag 3.35$$
Let us estimate the corresponding $|F(z)|$
 and $|G(z)|$ on the unit circle $|z|=1.$
 Using (3.35) in
 (2.8) we get
$$
|F(z)|\big|_{|z|=1}=|V_b|^b.\tag 3.36$$
On the other hand, using (3.35) in (2.9) we obtain
$$
|G(z)|\big|_{|z|=1}\le c\, |V_b|^{b-1}, \tag 3.37$$
where we have used the fact that $G(z)$ can be viewed as a finite
 sum of monomials in $(V_1,\dots, V_b)$ of degree $b-1$ or less
 multiplied with some nonnegative integer power of $|z|.$
Thus, (2.9) implies that each such monomial is bounded
 by a constant multiple of $|V_b|^{b-1},$ and since there are only a finite number of such
 monomials, there exists a positive constant $c$ depending on $b$
 for which (3.37) holds.
We can choose $|V_b|$ large enough so that $|V_b|>c$, and
 hence from (3.36) and (3.37) we get
$$
|G(z)|\big|_{|z|<1}<|F(z)|\big|_{|z|=1}. \tag 3.38$$
 Using (3.38) in the decomposition (2.7), we can
 apply Rouch\'e's theorem of complex variables and conclude that the
 number of zeros of $f_0(z)$ inside the unit circle $|z|=1$
 coincides with the number of the zeros of $F(z)$ inside that unit
 circle. By Theorem~2.1(d), the number of zeros of $f_0(z)$
 inside the unit circle is equal to $N,$ the number of bound states.
 On the other hand, using (3.35) in (2.8) we have
$$
|F(z)|=|V_b|^b\, |z|^b, \tag 3.39$$
and hence $F(z)$ has exactly $b$ zeros inside the unit circle, and
in fact all such zeros occur at $z=0.$ Thus, we have proved that
there exists at least one potential
with $V_b\ne 0$ in class $\Cal A_b$ for which $N=b.$
In fact, since we can choose $V_b$ in an infinite number of ways such that
 (3.35) and $|V_b|>c$ are satisfied, it follows that there
 are indeed an infinite number of potentials in class $\Cal A_b$ for
 which we have $N=b.$
 \qed

The next theorem shows that for each integer $k$ in the set $\{0,1,\dots,b\}$
there are an infinite
number of potentials in class $\Cal A_b$ having exactly
$k$ bound states. We recall that $V_b\ne 0$
for each potential in $\Cal A_b.$

\noindent {\bf Theorem 3.6} {\it For any fixed $b$
with $b\ge 1$ and for each nonnegative
integer $k$ in the set
$\{0,1,\dots,b\},$ there correspond infinitely many potentials
in class $\Cal A_b$ having exactly $k$ bound states.}

\noindent PROOF: The proof for $k=0$
is given in Theorem~3.4 and the proof for $k=b$ is given in
Theorem~3.5. So, it is enough to provide the proof
for $k$ in the set $\{1,\dots,b-1\}.$
In the proof of Theorem~3.5, the potentials explicitly
constructed in class $\Cal A_b$ with $b$ bound states
and with $V_b\ne 0$ are all generic, i.e. the corresponding
Jost functions $f_0(z)$ do not vanish at $z=\pm 1.$ This can be seen
 from (2.7) and (3.38) by noting that on the unit circle
 $|z|=1$ we have
$$|f_0(z)|\big|_{|z|=1}=|F(z)+G(z)|\big| _{|z|=1}\ge |F(z)|\big| _{|z|=1}-|G(z)|\big| _{|z|=1}>0,$$
and hence $f_0(z)$ cannot vanish on the unit circle and in particular
it cannot vanish at $z=\pm 1.$ Thus, for each integer $k$
in the set $\{1,\dots,b-1\}$ we conclude that there is at least one generic potential $V$ with exactly $k$ bound states in class $\Cal A_k$ with
$V_k\ne 0$ and $V_n=0$ for $n> k.$ Let us assume that
the corresponding bound-state zeros of the Jost function
$f_0(z)$ occur at $z=z_l$ with $l=1,\dots,k.$
We know from Theorem~2.1(d) that
each such $z_l$ is simple and belongs to
the set $z\in(-1,0)\cup(0,1).$ Let us continuously perturb
the potential $V$ to $\tilde V$ in such a way that
$\tilde V_n=V_n$ for $n\le k,$ $\tilde V_n=\epsilon$
for $k<n\le b,$ and
$\tilde V_n=0$ for $n>b,$ where
$\epsilon$ is a nonzero real parameter. The perturbed potential
$\tilde V$ belongs to class $\Cal A_b,$
and let us use $\tilde f_0(z)$
for the corresponding Jost function.
By Theorem~2.1(b) we know that as we perturb
$V$ to $\tilde V$ continuously, the coefficients
of $f_0(z)$ change continuously, and hence the zeros
of $f_0(z)$ also change continuously.
We claim that when $\epsilon$ is small
the number of zeros of $\tilde f_0(z)$ in
$z\in(-1,0)\cup(0,1)$ must be equal to $k.$
This can be seen as follows.
Because of Theorem~2.1(g), as we introduce the perturbation
none of
$z_l$-values can abruptly change to nonreal values,
and because of Theorem~2.1(c) none of those
$z_l$-values can change to zero.
Thus, the only way to change the number of
bound states during the continuous perturbation
would be for a zero of $f_0(z)$ moving into or out of
$z\in(-1,0)\cup(0,1)$ through $z=-1$ or $z=1.$
By choosing $\epsilon$ small enough, we know that
we must have $\tilde f_0(\pm 1)\ne 0$
because we have $f_0(\pm 1)\ne 0.$
Thus, we have shown that for small enough
$\epsilon$ we have
$\tilde V$ belonging to class $\Cal A_b$
with $\tilde V_b\ne 0$ and
the corresponding perturbed
operator has exactly $k$ bound states.
Let us remark that, as we continuously perturb the
potential $V$ to $\tilde V,$ the degree of
$f_0(z)$ abruptly changes from $2\,k-1$
to $2\,b-1,$ which is the degree of
$\tilde f_0(z).$ However, the additional
zeros of the Jost function enter the complex-$z$ plane
 from $z=\infty$ and hence
 they do not increase the number of bound states.
Thus, our proof is complete. \qed

Having proved one of our main results in Theorem~3.6, in the
next theorem we prove our other main result.

\noindent {\bf Theorem 3.7} {\it Assume that the potential $V$ appearing in (1.1)
 belongs to class $\Cal A_b$ for some positive integer $b.$
 Then, the number of bound states, denoted by $N$,
 for the corresponding discrete Schr\"odinger operator
 associated with (1.1) and (1.2)
 satisfies the
 inequality}
$$0\le N\le b.\tag 3.40$$
{\it We remark that the result trivially holds also when
$b=0.$}

\noindent PROOF:
Let us first indicate that (3.40) holds
when $b=0$ because in that case
$\Cal A_b$ consists of the trivial potential $V_n\equiv 0$ for
which we already know [2] that $N=0.$
By Theorem~2.1(d) we have
$$Z(-1,0)+Z(0,1)=N,\tag 3.41$$
where we recall that $Z(-1,0)$ and $Z(0,1)$ denote the number
 of zeros of the Jost function $f_0(z)$ in the respective intervals
 $z\in (-1,0)$ and $z\in(0,1)$.
 With the help of (3.41), from (3.18)
  and (3.19) we conclude that
$$Z(-\infty,-1]+Z[1,+\infty)\ge\cases
   0,\qquad N=0, \\
   N-1,\qquad N=1, \\
   N-2,\qquad N\geq 2.\endcases\tag 3.42$$
  Using (3.41) and (3.42) in (2.13), we obtain
$$2b-1\ge N+2\,Z_c+
\cases
   0,\qquad N=0, \\
   N-1,\qquad N=1, \\
   N-2,\qquad N\geq 2.\endcases\tag 3.43$$
Since $Z_c\ge 0,$ from (3.43) we get
 $$2b-1\ge
\cases
   N,\qquad N=0, \\
   2\,N-1,\qquad N=1, \\
   2\,N-2,\qquad N\geq 2,\endcases\tag 3.44$$
 or equivalently
  $$2b-1\ge
\cases
   0,\qquad N=0, \\
   1,\qquad N=1, \\
   2\,N-2,\qquad N\geq 2.\endcases\tag 3.45$$
  From (3.45) we see that
$$b\ge
  \cases
   \ds\frac{1}{2},\qquad N=0, \\
   1,\qquad N=1, \\
   N-\ds\frac{1}{2},\qquad N\geq 2.\endcases\tag 3.46$$
  From (3.46) we infer that
 $$N\le
  \cases
   b,\qquad N=0, \\
   b,\qquad N=1, \\
   b+\ds\frac{1}{2},\qquad N\geq 2.\endcases\tag 3.47$$
     From the third line of (3.47) we conclude that any
  potential in class $\Cal A_b$ satisfies
$N<b$ or $N\le b.$
    By Theorem~3.5 we know that $N\le b$ holds
  for an infinite number of potentials in class $\Cal A_b$. Hence
  we must have $N\le b$ for any potential in class $A_b.$ \qed

\vskip 10 pt
\noindent {\bf 4. SOME EXPLICIT EXAMPLES}
\vskip 3 pt

In this section we illustrate the results presented
in Section~3 with some
explicit examples.

In our first example we consider all the
 possibilities for the bound states
 for potentials in class $\Cal A_b$ with
$b=1.$

\noindent {\bf Example 4.1} Consider the compactly-supported potentials
$V_n$ in class $\Cal A_b$ with $b=1,$ and hence we have $V_n=0$ for $n\ge 2$ and $V_1\ne 0.$
 From (2.14) we know that the corresponding
Jost function is given by
$$f_0=1+V_1 z.\tag 4.1$$
According to Theorem~3.7 we must have $0\le N\le 1$
for the number of bound states, and
by Theorem~3.6 we must have $N=0$ for an infinite number of $V_1$-values
  and we must have $N=1$ for an infinite number of $V_1$-values.
 From (4.1) we see that the only zero of $f_0(z)$ occurs at
 $z=\alpha_1$ with $\alpha_1=-1/V_1.$ Thus, when $0<|V_1|<1$
 we have $|\alpha_1|>1$ and hence $\alpha_1\not\in (-1,0)\cup(0,1),$
 indicating that $N=0.$ On the other hand, when $|V_1|>1,$
 we have $0<|\alpha_1|<1$ and hence $\alpha_1\in (-1,0)\cup(0,1),$
 indicating that $N=1.$ Let us also remark that the inequalities given
 is (3.18) and (3.19) hold when $N=0$ and
 $N=1.$

In the next example, we explore all the possibilities for the bound
 states in class $\Cal A_b$ when $b=2.$

\noindent {\bf Example 4.2} Consider the potential class $\Cal A_b$
with $b=2,$ and hence we have $V_n=0$ for $n>2$ and
$V_2\ne 0.$ From (2.14) we obtain the
 corresponding Jost function as
$$f_0(z)=1+\left(V_1+V_2\right)\,z+V_1\,V_2\,z^2+V_2\,z^3.\tag 4.2$$
 In this case $f_0(z)$ has three zeros $\alpha_1,$ $\alpha_2,$
 $\alpha_3$ and we have
$$
 f_0(z)=\left(1-\frac{z}{\alpha_1}\right)
 \left(1-\frac{z}{\alpha_2}\right)
 \left(1-\frac{z}{\alpha_3}\right). \tag 4.3$$
 Equating the corresponding coefficients in (4.2) and (4.3)
 we get
 $$\cases
   V_1+V_2 =-\ds\frac{1}{\alpha_1}-\ds\frac{1}{\alpha_2}-\ds\frac{1}{\alpha_3},\\
   V_1\,V_2 =\ds\frac{1}{\alpha_1\,\alpha_2}
   +\ds\frac{1}{\alpha_1\,\alpha_3}+\ds\frac{1}{\alpha_2\,\alpha_3},\\
   V_2= -\ds\frac{1}{\alpha_1\,\alpha_2\,\alpha_3}.\endcases\tag 4.4$$
 The nonlinear algebraic equations given in (4.2) impose certain restrictions on
 the locations of $\alpha_1,$ $\alpha_2,$ $\alpha_3$ in the
 complex-$z$ plane when $V_1$ and $V_2$ take real values. We can
 view $V_1$ and $V_2$ as the two solutions to the quadratic equation
$$x^2-\left(V_1+V_2\right)\,x+V_1\,V_2=0,\tag 4.5$$
 where the solutions are given by
$$x=\ds\frac{V_1+V_2\pm \ds\sqrt{(V_1+V_2)^2-4V_1\,V_2}}{2}.\tag 4.6$$
 Thus, $V_2$ must be equal to one of the two quantities on the
 right-hand side of (4.6).
  Using the first two lines of (4.4) in (4.5) we get
$$x=
\ds \frac{-\left(\ds\frac{1}{\alpha_1}+\ds\frac{1}{\alpha_2}+\ds\frac{1}{\alpha_3}
 \right)\pm
 \ds\sqrt{\left(\ds\frac{1}{\alpha_1}+\ds\frac{1}{\alpha_2}+\ds\frac{1}{\alpha_3}
 \right)^2-4\left(\frac{1}{\alpha_1\,\alpha_2}
   +\frac{1}{\alpha_1\,\alpha_3}+\frac{1}{\alpha_2\,\alpha_3}
 \right)}}{2}.\tag 4.7$$
 Then comparing (4.7) with the third line of (4.4) we
see that we must have
$$
 -\ds\frac{2}{\alpha_1\,\alpha_2\,\alpha_3}+
 \left(\ds\frac{1}{\alpha_1}+\ds\frac{1}{\alpha_2}+\ds\frac{1}{\alpha_3}
 \right)=
 \sqrt{\left(\ds\frac{1}{\alpha_1}+\ds\frac{1}{\alpha_2}+\ds\frac{1}{\alpha_3}
 \right)^2-4\left(\ds\frac{1}{\alpha_1\,\alpha_2}
   +\ds\frac{1}{\alpha_1\,\alpha_3}+\ds\frac{1}{\alpha_2\,\alpha_3}
 \right)},\tag 4.8$$
or
$$
 -\ds\frac{2}{\alpha_1\,\alpha_2\,\alpha_3}+
 \left(\ds\frac{1}{\alpha_1}+\ds\frac{1}{\alpha_2}+\ds\frac{1}{\alpha_3}
 \right)=-
 \sqrt{\left(\ds\frac{1}{\alpha_1}+\ds\frac{1}{\alpha_2}+\ds\frac{1}{\alpha_3}
 \right)^2-4\left(\ds\frac{1}{\alpha_1\,\alpha_2}
   +\ds\frac{1}{\alpha_1\,\alpha_3}+\ds\frac{1}{\alpha_2\,\alpha_3}
 \right)},\tag 4.9$$
 In order for (4.8) and (4.9) to hold we must have
$$\left(\alpha_1\,\alpha_2\,\alpha_3\right) \left(\alpha_1+\alpha_2+\alpha_3\right)+\left(\alpha_1\,\alpha_2
+\alpha_1\,\alpha_3+\alpha_2\,\alpha_3\right)=1,
\tag 4.10$$
which is obtained by squaring both sides of (4.8) and (4.9) and then
simplifying the resulting expressions.
The conditions (4.8) and (4.9) impose various
 restrictions  on $\alpha_1,$ $\alpha_2,$ and $\alpha_3.$ For example,
 we cannot have $\alpha_1,$ $\alpha_2,$ $\alpha_3$ all located in the
  interval $[1,+\infty).$ Otherwise, the left-hand side of
  (4.10) would be greater than one. Similarly, we cannot have $\alpha_1,$ $\alpha_2,$ $\alpha_3$ all
located in the interval $(-\infty,-1].$ Otherwise, the left-hand
sides of (4.10) would again be greater than one.
Similarly, we cannot have $\alpha_1\ge 1$ while
$\alpha_3=\alpha_2^*$ with $\text{Re}\,[\alpha_3]\ge 1$
because that would make the left-hand side
greater than one.
In a
similar way we cannot have $\alpha_1\in (-1,0)$ while
$\alpha_3=\alpha_2^*$ with $\text{Re}\,[\alpha_3]\ge 1$
because that would again
make the left-hand side of (4.10)
greater than one.
Similarly,
we cannot have $\alpha_1\in(0,1)$ while $\alpha_3=\alpha_2^*$
with $\text{Re}\,[\alpha_3]\le -1.$ On the other hand, for example, a double real
resonance and a bound state is possible with
$$\alpha_1=\alpha_2=2,\quad \alpha_3=-\frac{3}{2}+\sqrt{3}=
0.23205\overline{1},\tag 4.11$$
 which
  correspond to
  $$V_1=-\frac{5}{2}+\sqrt{3}, \quad
V_2=-\frac{1}{2}+\frac{1}{\sqrt{3}}.
\tag 4.12$$
Here, we use an overline on a digit to indicate a round off.
 For example, a double real resonance with another real resonance is
 possible with
$$\alpha_1=\alpha_2=2, \quad \alpha_3=-\ds\frac{3}{2}-\sqrt{3}=-3.23205\overline{1},
$$
 with the potential values
$$V_1=-\frac{5}{2}+\sqrt{3},\quad
V_2=-\frac{1}{2}+\frac{1}{\sqrt{3}}.$$
We also get a double real
resonance and a bound state with
$$\alpha_1=\alpha_2=-2,\quad \alpha_3=\frac{3}{2}-\sqrt{3}=
-0.23205\overline{1},\tag 4.13$$
 which
  correspond to
  $$V_1=\frac{5}{2}-\sqrt{3}, \quad
V_2=\frac{1}{2}-\frac{1}{\sqrt{3}}.
\tag 4.14$$
The restriction (3.18) indicates that if we have two bound
  states with $\alpha_1$ and $\alpha_2$
  both being in the interval $(-1,0),$ then we must have a
  real resonance with $\alpha_3\le -1.$ Similarly, the restriction
  (3.19) indicates that if we have two bound states with
   $\alpha_1$ and $\alpha_2$ both being in
   the interval $z\in(0,1)$, then we must have a real resonance
   with $\alpha_3\geq1$. Let us remark that we can have two bound
   states and one real resonance by choosing $\alpha_1,\alpha_2,\alpha_3$
   appropriately so that the corresponding $V_1$ and $V_2$ are
   real valued. For example, for
$$V_1=-\sqrt{5},\quad V_2=\frac{4}{\sqrt{5}},$$
   we get
$$\alpha_1=\frac{1}{2},\quad \alpha_2=-\frac{1}{2},\quad \alpha_3=\sqrt{5}.
$$
Choosing
$$V_1=\sqrt{5},\quad V_2=-\frac{4}{\sqrt{5}},$$
   we get
$$\alpha_1=\frac{1}{2},\quad \alpha_2=-\frac{1}{2},\quad \alpha_3=-\sqrt{5}.$$
 Choosing
$$V_1=\frac{-13+\sqrt{22}}{3},\quad V_2=-4-4\sqrt{\frac{2}{11}},$$
we get
$$\alpha_1=\frac{1}{6},\quad \alpha_2=\frac{1}{2},\quad \alpha_3=
 \frac{11-\sqrt{22}}{3}=2.1031\overline{9},$$
  and choosing
$$V_1=\frac{-13-\sqrt{22}}{3},\quad V_2=4+4\sqrt{\frac{2}{11}},$$
 we get
$$\alpha_1=-\frac{1}{6},\quad \alpha_2=-\frac{1}{2},\quad \alpha_3=
 \frac{-11+\sqrt{22}}{3}=-2.1031\overline{9}.$$

In the following example, we illustrate
some of the possibilities for the number of bound states
and resonances for potentials in
class $\Cal A_b$ with $b=3.$

\noindent {\bf Example 4.3} Consider the potential class $\Cal A_b$
with $b=3,$ and hence $V_n=0$ for $n>3$ and $V_3\ne 0.$ From (2.14)
we see that the corresponding Jost function $f_0(z)$
 is expressed in terms of $V_1,$ $V_2,$ $V_3$ as
$$\aligned
f_0(z)=1+&\left(V_1+V_2+V_3\right)z+\left[V_1\,V_2+(V_1+V_2)\,V_3\right]
z^2\\
&+
  \left[V_2+V_3(1+V_1\,V_2)\right]z^3
  +V_3\left(V_1+V_2\right)z^4+V_3\,z^5.
  \endaligned \tag 4.15$$
In terms of the zeros $\alpha_1,$ $\alpha_2,$ $\alpha_3,$
$\alpha_4,$ $\alpha_5$
  of $f_0(z)$ we have representation
$$
  f_0(z)=\left(1-\frac{z}{\alpha_1}\right)\left(1-\frac{z}{\alpha_2}\right)\left(1-\frac{z}{\alpha_3}\right)
  \left(1-\frac{z}{\alpha_4}\right)\left(1-\frac{z}{\alpha_5}\right). \tag 4.16$$
  By equating the corresponding coefficients in (4.15)
  and (4.16) we express
 $\alpha_1,$
 $\alpha_2,$ $\alpha_3,$ $\alpha_4,$ $\alpha_5$
 in terms of $V_1,$ $V_2,$ $V_3$ as a nonlinear system of five equations
 given by
 $$\cases
  V_1+V_2+V_3 =-\left(\ds\frac{1}{\alpha_1}+\ds\frac{1}{\alpha_2}+
  \ds\frac{1}{\alpha_3}+\ds\frac{1}{\alpha_4}+\ds\frac{1}{\alpha_5} \right),\\
  \stretch
  V_1\,V_2+\left(V_1+V_2\right)\,V_3
 = \ds\frac{1}{\alpha_1\alpha_2}+\ds\frac{1}{\alpha_1\alpha_3}+\cdots+
  \ds\frac{1}{\alpha_4\alpha_5},\\
  \stretch
  V_2+V_3\left(1+V_1\,V_2\right)=-\left(\ds\frac{1}{\alpha_1\,\alpha_2\,\alpha_3}+
\ds\frac{1}{\alpha_1\,\alpha_2\,\alpha_4}+\cdots+
 \ds\frac{1}{\alpha_3\,\alpha_4\,\alpha_5}
  \right), \\
  \stretch
 \left(V_1+V_2\right)\,V_3=\ds\frac{1}{\alpha_1\,\alpha_2\,\alpha_3\,\alpha_4}+
 \frac{1}{\alpha_1\,\alpha_2\,\alpha_3\,\alpha_5}+\cdots
+\ds\frac{1}{\alpha_2\,\alpha_3\,\alpha_4\,\alpha_5},\\
\stretch
  V_3=-\ds\frac{1}{\alpha_1\,\alpha_2\,\alpha_3\,\alpha_4\,\ds\alpha_5}
.\endcases\tag 4.17$$
Notice that if (4.17) has a solution, then we can change
 the sign of each of $V_1,$
 $V_2,$
 $V_3$ and $\alpha_1,$
 $\alpha_2,$ $\alpha_3,$
 $\alpha_4,$
$\alpha_5$ and get another solution. The nonlinear relations given in (4.17)
 put certain restrictions on the locations of $\alpha_1,$
 $\alpha_2,$ $\alpha_3,$ $\alpha_4,$
 $\alpha_5$ on the complex-$z$ plane in order to have $V_1,$ $V_2,$ $V_3$
as real-valued
constants. The system in (4.17) can be solved to express
$V_1,$ $V_2,$ $V_3,$ $\alpha_5$
 in terms of $\alpha_1,$ $\alpha_2,$ $\alpha_3,$ $\alpha_4$ by solving the
 fifth line in (4.17) for $V_3,$ then solving the fourth line
 for $V_2,$ then solving the first line for $\alpha_5,$ and
 solving the second line for $V_1.$ Then, we can use the resulting
 expressions for $V_1,$ $V_2,$ $V_3,$ $\alpha_5$ in the third line of (4.17)
 to get the consistency. We then obtain a consistency equation involving
$\alpha_1,$ $\alpha_2,$ $\alpha_3,$ $\alpha_4.$ By assigning various
allowable values for $\alpha_1,$ $\alpha_2,$ $\alpha_3,$ $\alpha_4,$ we can
then produce some explicit examples. For example, by choosing
$$\alpha_2=\alpha_1,\quad
\alpha_3=\alpha_4=\alpha_1^\ast,$$
we can
demonstrate the existence of a double nonreal resonance with
$\alpha_1=-0.3196\overline{8}+2\,i$ corresponding to
$$\alpha_5=-0.60017\overline{2},\quad
V_1=1.1327\overline{9},\quad V_2=0.74610\overline{6},\quad
V_3=0.099012\overline{9},$$
 and we
 observe that this case has exactly one bound state at $z=\alpha_5$.
  We obtain another example with
$$\alpha_1=1.161\overline{3}+i,
\quad \alpha_2=\alpha_1,\quad
\alpha_3=\alpha_4=\alpha_1^\ast,\quad
\alpha_5=0.2779\overline{7},\tag 4.18$$
$$V_1=-1.8911\overline{4},\quad V_2=-3.0320\overline{2},\quad
V_3=-0.652\overline{2},\tag 4.19$$
which indicates that we have one bound state
at $z=\alpha_5,$ a double
complex resonance at $z=\alpha_1,$ and a
double complex resonance at $z=\alpha_1^\ast.$

In the final example below we present a specific example where $N=b$
is attained in class $\Cal A_b.$

\noindent {\bf Example 4.4}
Let us choose the potential appearing in (1.1) as
$$
V_n=\cases (-1)^n\,2,\qquad 1\le n\le b,\\
\stretch
0,\qquad n>b.\endcases\tag 4.20$$
so that it belongs to class $\Cal A_b.$
 Using (2.14) we then get the Jost function $f_0(z)$
 explicitly expressed as a polynomial in $z$ of degree $2b-1.$
Using the symbolic computing system Mathematica,
 we evaluate the zeros of $f_0(z)$ numerically and observe
 that, e.g. for each $b=1,2,\dots, 110$ the resulting $f_0(z)$
 has exactly $b$ real zeros in $z\in(-1,0)\cup(0,1).$
  This numerically confirms the result presented in Theorem 3.5.
  We remark that as $b$ increases some of the zeros of $f_0(z)$
  start getting closer to $z=\pm 1.$ In that case, one needs to increase
  the accuracy of the numerical program used to evaluate
  the zeros of a polynomial function to avoid any discrepancies. If one replaces
  the value of $2$ in the first line of (4.20) with a larger value,
  then the zeros of $f_0(z)$ in the set
  $(-1,0)\cup(0,1)$ move away from $z=\pm 1$ and hence
  it becomes easier to confirm
  $N=b$ for large $b$-values during the
  numerical evaluation of the zeros of
  $f_0(z).$

\vskip 5 pt

\noindent {\bf Acknowledgments.}
The first author expresses his gratitude to the Institute of Physics and Mathematics of the Universidad Michoacana de San Nicol\'as de
Hidalgo, M\'exico for its hospitality.
The second author was partially supported by SNI-CONACYT and CIC-UMNSH, Mexico.

\vskip 5 pt

\noindent {\bf{References}}

\item{[1]} T. Aktosun, A. E. Choque-Rivero, and V. G. Papanicolaou, {\it Darboux transformation for the discrete Schr\"odinger equation,} preprint.

\item{[2]}
T. Aktosun and V. G. Papanicolaou, {\it Inverse problem with transmission eigenvalues for the discrete Schr\"odinger equation,}
J. Math. Phys. {\bf 56}, 082101 (2015).

\item{[3]}
K. M. Case and M. Kac,
{\it A discrete version of the inverse scattering problem,}
J. Math. Phys. {\bf 14}, 594--603 (1973).

\item{[4]} D. Damanik and G. Teschl, {\it
Bound states of discrete Schr\"odinger operators with super-critical inverse square potentials,}
Proc. Amer. Math. Soc. {\bf 135}, 1123--1127 (2007).

\item{[5]}
G. Teschl,
{\it Jacobi operators and completely integrable nonlinear lattices,}
Amer. Math. Soc.,  Providence, RI, 2000.

\item{[6]} B. N. Zakhariev  and A. A. Suzko, {\it
Direct and inverse problems,} Springer-Verlag, Berlin, 1990.

\end